\documentclass[letterpaper, 10 pt, conference]{ieeeconf}  % Comment this line out if you need a4paper
\usepackage{graphicx}
\usepackage{amsmath}
\usepackage{subcaption}
\usepackage{caption}
\usepackage{makecell}

\IEEEoverridecommandlockouts                              % This command is only needed if 
\title{\LARGE \bf
Eye Movement Feature-Guided Signal De-Drifting in Electrooculography Systems
}

\author{Lianming Hu, Xiaotong Zhang and Kamal Youcef-Toumi% <-this % stops a space
\thanks{All authors are with the Mechatronics Research Laboratory, Massachusetts Institute of Technology, Cambridge, MA, 02139, USA
        {\tt\small \{lianmhu, kevxt, youcef\}@mit.edu }}%
\thanks{This research was made possible by the support and partnership
of King Abudlaziz City for Science and Technology
(KACST) through the Center for Complex Engineering
Systems at Massachusetts Institute of Technology (MIT) and
KACST.}%
}

\begin{document}

\maketitle
\thispagestyle{empty}
\pagestyle{empty}

%%%%%%%%%%%%%%%%%%%%%%%%%%%%%%%%%%%%%%%%%%%%%%%%%%%%%%%%%%%%%%%%%%%%%%%%%%%%%%%%
\begin{abstract}

Electrooculography (EOG) is widely used for gaze tracking in Human-Robot Collaboration (HRC). However, baseline drift caused by low-frequency noise significantly impacts the accuracy of EOG signals, creating challenges for further sensor fusion. This paper presents an Eye Movement Feature-Guided De-drift (FGD) method for mitigating drift artifacts in EOG signals. The proposed approach leverages active eye-movement feature recognition to reconstruct the feature-extracted EOG baseline and adaptively correct signal drift while preserving the morphological integrity of the EOG waveform. The FGD is evaluated using both simulation data and real-world data, achieving a significant reduction in mean error. The average error is reduced to 0.896° in simulation, representing a 36.29\% decrease, and to 1.033° in real-world data, corresponding to a 26.53\% reduction. Despite additional and unpredictable noise in real-world data, the proposed method consistently outperforms conventional de-drifting techniques, demonstrating its effectiveness in practical applications such as enhancing human performance augmentation.

\end{abstract}

%%%%%%%%%%%%%%%%%%%%%%%%%%%%%%%%%%%%%%%%%%%%%%%%%%%%%%%%%%%%%%%%%%%%%%%%%%%%%%%%
\section{INTRODUCTION}

Human-robot collaboration (HRC) has become a trending solution in both industry and daily life scenarios. In some cases, it has also proven to be a better solution than traditional robotic systems, as it combines the strengths of both humans and robots: humans exhibit superior flexibility and adaptability, while robots excel in accuracy, cost-effectiveness, and efficiency \cite{10610657} \cite{kothari2023enhancedhumanrobotcollaborationusing}. However, to ensure an efficient and productive workflow in HRC, robust and effective communication between humans and robots must be established.  

As naturally efficient collaborators in HRC, humans have various ways to convey information, either explicitly or implicitly \cite{doi:10.1142/S0219843608001303}. Among all modalities of information transfer, the gaze is an implicit yet rich source of data that enables robots to infer human states, revealing both physical and cognitive activities, including attention, intention, and relevance \cite{zhang2024relevancedrivendecisionmakingsafer,zhang2024relevancehumanrobotcollaboration,zhang2025thesis}. This information is particularly valuable in enhancing task execution, planning, and prediction in HRC.

Currently, there are various ways of tracking the human gaze. A common method for gaze tracking nowadays is a camera-based oculography system that tracks certain parts of the eye’s movement relative to the reference point of the eye to predict the gaze \cite{MORIMOTO20054}. However, certain limitations are also found in the camera-based eye gaze tracking method. For example, a stable environment with an ideal level of illumination and a high-resolution camera is required to ensure the necessary eye image or video can be well captured. Also, this method cannot work if the user turns to the other direction where the camera fails to capture the eye region \cite{hernandezcruz2024bayesianintentionenhancedhuman}. All these limitations make camera-based eye-tracking expensive and cumbersome for a more dynamic environment such as HRC.

Electrooculography (EOG) is another gaze-tracking technology that utilizes sensors to detect changes in corneal-retinal potential (CRP) caused by human eye movement. It relies on tiny skin-surface electrodes as the only input, making it a simpler setup for many user-controlled devices while offering greater flexibility, as users can move freely. Additionally, since EOG does not require a camera, it avoids high computational resource demands, large power supply requirements, and privacy concerns. As a result, it has the potential to become a more practical and accessible gaze tracking technique in the HRC field.

Current EOG solutions still have limitations, and many challenges remain active topics of research today. One major challenge for EOG to be used in a long-term timeframe is the drift of the signal. The drift will slowly add error to the baseline of the EOG signal, causing a shifting of the user’s absolute gaze angle prediction from the true value. Much work has been done to address different solutions in mitigating the drift effect, including high pass filtering \cite{HPacuna2014eye}, polynomial fitting \cite{POLYhuda2015recognition}, signal differencing \cite{DIFFryu2019eog}, wavelet decomposition \cite{bulling2010eye}, and techniques based on baseline component recognition \cite{barbara2021temod}. These techniques are common methods for achieving de-drifting. However, each has its limitations, affecting practicality in different circumstances. For example, while high-pass filtering is a simple and effective solution for both offline and real-time de-drifting, it fails to preserve the original morphology of the EOG signal, distorting the waveform and altering the amplitude of key features such as saccades. In contrast, more complex systems, such as baseline component recognition, can successfully de-drift while preserving signal morphology but require prior knowledge of each subject’s cue-target information during signal acquisition. This necessitates a fixed trial protocol, which must be predefined and known by the system before data processing, significantly limiting its practicality in free-movement EOG use cases.

In this work, a novel Feature-Guided De-drifting (FGD) method was presented to minimize the drift effect. The FGD method extracts the drift trend from the raw EOG signal by first identifying saccades, one of the most critical features of eye movement. It then applies an adaptive window to accurately exclude saccades from the EOG signal, reconstruct the remaining signal as a new baseline, and finally use 1D multilevel wavelet decomposition to approximate the drift trend for de-drifting. Notably, while both FGD and the baseline component recognition system by N. Barbara et al. \cite{barbara2021temod} perform de-drifting through feature extraction on the EOG baseline, FGD does not require prior knowledge from a fixed trial protocol. This makes it well-suited for most HRC scenarios, where the user’s gaze angle and gaze transitions are not predetermined but freely chosen. Additionally, as a feature-guided approach, the proposed method has lower sampling frequency requirements, making it more suitable for cost-effective systems with greater practicality. In both simulation and real data testing, this method has successfully removed the drift without distorting the morphology of the EOG signal.

More specifically, the key contributions of this paper are as follows:

\begin{itemize}
    \item A novel de-drifting approach for EOG signals in free-movement use cases, leveraging active feature extraction to enhance de-drifting while preserving the morphology of the original signal.
    \item An adaptive feature extraction method that dynamically aligns with the actual saccade time window while reconstructing the baseline, improving the de-drifting process's precision, robustness, and signal stability.
    \item A seamless transition from simulation to real-world implementation, demonstrating the proposed method’s effectiveness in preserving EOG signal morphology while achieving accurate de-drifting.
\end{itemize}

\section{RELATED WORK}

To the best of our knowledge, our work is very unique. In the following sub-sections, some works related to ours are summarized.

\subsection{Human Eye Gaze Tracking}

The primary patterns of human eye movement can be categorized into three types: fixation, blink, and saccade \cite{bulling2010eye}. Various technologies have been developed to track eye movements and predict gaze direction. For instance, contact-based tracking methods require physical contact with the user such as contact lenses, electrodes, head-mounted devices, etc. In contrast, contactless eye trackers have also been developed to minimize intrusiveness by tracking eye movement without physical contact. These systems are mostly vision-based, with camera-based eye tracking being the most common approach \cite{MORIMOTO20054}.

\subsubsection{Camera-Based Eye Tracking}

Camera-based eye tracking requires the camera to monitor the features of the human eye and perform computer vision-related analyses of the figure or video captured. For example, pupil-corneal reflection techniques use a light source to create a reference point, then detect the pupil’s center position, and calculate the gaze based on that. This technique has been shown to be accurate and simple for experiment setups \cite{5601753} \cite{6525411}, thus implemented by many camera-based eye trackers. The latest camera-based gaze detection is usually a complex prediction system that also contains multiple facial behavior analysis subsystems, for example, the OpenFace model by T. Baltrušaitis et al. \cite{baltruvsaitis2016openface}, and the PtGaze system by X. Zhang et al. \cite{zhang2020eth}. 

%In this work, after testing various camera-based gaze tracking systems, we decided to implement the PtGaze system mentioned above as our ground truth baseline for its accuracy and compatibility with our system. 

\subsubsection{EOG-Based Eye Tracking}

The EOG-based system uses sensors to measure the electrical signal caused by human eye movement. In the human eye, the cornea, located at the front, carries a positive charge, while the retina, positioned at the back, is negatively charged. As a result, the eye can be modeled as a dipole, with its orientation varying in accordance with the optical axis. It is worth noting that the difference of potential between cornea and retina, referred to as the corneo-retinal potential (CRP), is assumed to be constant in the range of 0.4–1.0 mV depending on the subject \cite{EOGbasic} \cite{brown2006iscev}. As a result, the EOG technology can monitor and record the potential differences, analyzing how the potential differences change as humans move their eyes and thus calculate human gaze angle using these data. 

\subsection{EOG Drift}

The EOG signal is affected by superimposed low-frequency noise that is unrelated to eye movement and can exhibit either linear or nonlinear characteristics. The source of drift can be from background noise, electrode polarization, illumination level change, the pressure and resistance change of the contact area on the skin, etc \cite{barbara2020driftcomparison} \cite{bulling2010eye}. Drift primarily affects the EOG signal baseline by gradually introducing errors into the predicted gaze angle. However, high-frequency eye movements, such as saccades, remain unaffected by drift due to their short duration \cite{barbara2020driftcomparison} \cite{manabe2015direct}. 

The EOG signal $E(t)$ can be decomposed into three components: the eye movement generated signal $f(t)$, the baseline drift $d(t)$, and noise $w(t)$. In this model, baseline drift is treated separately from other noise, as this study focuses on removing its effects. The model is expressed as follows:

\begin{equation}
    E(t) = f(t) + d(t) + w(t) 
\end{equation}

Many techniques have been used to minimize the drift effect $d(t)$ so the de-drifted signal $E’ (t)$ can be obtained. In this work, some of the most popular and latest methods have been selected to compare against the FGD and will be reviewed in the following sections.

\subsubsection{Polynomial Fitting}

An approach to removing drift from the EOG signal is to estimate the drift component, \( \hat{d}(t) \), by fitting a polynomial function of n\textsuperscript{th} order to the drift trend \cite{pettersson2013algorithm}. However, polynomial fitting considers de-drifting only in the time domain, failing to capture the frequency-related characteristics of the drift effect. In contrast, our method not only preprocesses the signal in the time domain but also applies frequency-domain-based de-drifting to further mitigate this issue.

\subsubsection{High-pass Filtering}

High-pass filtering is also a popular method to be used to mitigate the drift on EOG signal. The cutoff frequency for the de-drift high pass filter is not commonly set and is often found to be less than 0.5HZ \cite{HPacuna2014eye}. Nevertheless, high-pass filtering tends to distort the EOG signal. In contrast, we leverage the 1D multilevel wavelet decomposition to preserve both time and frequency information. 

\subsubsection{1D Multilevel Wavelet Decomposition}

Another approach is through 1D multilevel wavelet decomposition. During the 1D multilevel wavelet decomposition process, the discrete wavelet transform is applied to the original EOG signal to decompose it and obtain approximation coefficients $A_j[n]$ and detail coefficients $D_j[n]$ at chosen level $j$. After $j$ levels of decomposition, the lowest frequency coefficient $A_J$ can then be used to reconstruct the approximation of drift trend \( \hat{d}(t) \) and used to de-drift the original signal. Meanwhile, this approach utilizes the entire original signal for de-drifting, failing to account for the impact of high-frequency eye movement. Instead, our system employs a Baseline Reconstructor, which mitigates these effects by isolating eye movement components, thereby achieving a more effective de-drifting outcome.

\section{METHODOLOGY}

\subsection{Signal Acquisition}

The EOG signal acquisition system utilizes conventional skin electrodes, specifically Ag/AgCl bio-sensors with gel-applied pads. Since the primary focus of this paper is on de-drifting the signal, a single-channel horizontal EOG (HEOG) system is used for simplicity. Three electrodes are placed on the subject’s head, as illustrated in Fig. 1. Electrodes 1 and 2 are placed on the outside of the eyes to capture horizontal eye movement-related signals, while electrode 3 serves as the reference node and is positioned on the forehead. With potential signal captured from electrodes 1 and 2, the $E(t)$ can be calculated through

\begin{figure}[!t]
\begin{center}
\includegraphics[width=0.4\columnwidth]{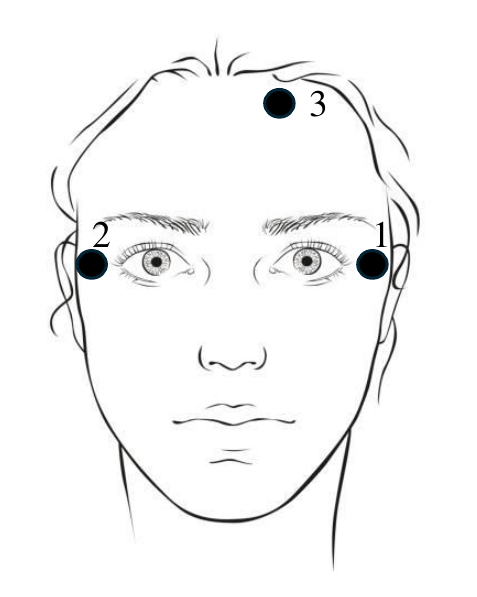}
\end{center}
\caption{EOG Electrodes setup configuration}
\label{fig: Electrode}
\end{figure}

\begin{equation}
    E(t) = V_2(t) - V_1(t)
\end{equation}

Where the $V_1(t)$ and $V_2(t)$ denote the voltage signal captured by electrode 1 and electrode 2, respectively. The EOG signal will first pass through a SRS SIM983 scaling amplifier. A gain of 300 is used in this stage. Then, a low-pass analog filter is used to filter out the high frequency noise with a cutoff frequency of 30HZ. An Arduino UNO is then used as an analog-to-digital converter (ADC) to pass the signal for digital processing. 

In the digital processing stage, a blink removal will be performed on the signal $E(t)$. The blink activity, when reflecting on the EOG signal captured, is shown as a short positive pulse on the baseline value that is mostly within a time window $< 400$ ms \cite{bulling2010eye}. Based on this condition, the algorithm detects and removes blink artifacts by identifying two consecutive, oppositely directed surges in the signal derivative that exceed the threshold value within a specified time window. As a result, a blink artifact-free EOG signal is obtained. For simplicity, we will denote $E(t)$ in the following sections as the blink artifact-free raw HEOG input.

\subsection{Feature Guided De-drifting}

In Fig. 2, the processing flow chart of the proposed method has been illustrated. The high-level structure can be broken down into two phases: (i) Feature extraction preprocessing and (ii) Drift trend approximation.

\begin{figure*}[!t]
\begin{center}
\includegraphics[width=2\columnwidth]{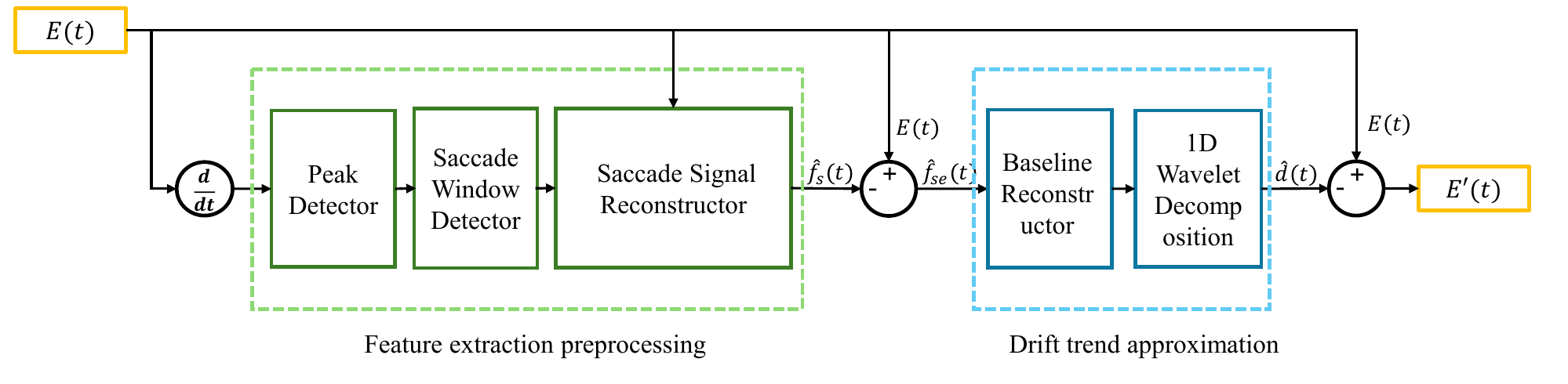}
\end{center}
\caption{Feature Guided De-drifting method flow}
\label{fig: Method}
\end{figure*}

\subsubsection{Feature extraction preprocessing}

%\begin{figure}[!t]
%\begin{center}
%\includegraphics[width=0.9\columnwidth]{Derivative_Peak.pdf}
%\end{center}
%\caption{Derivative plot for peak detection}
%\label{fig: coordinates}
%\end{figure}

%\begin{figure}[!t]
%\begin{center}
%\includegraphics[width=0.9\columnwidth]{Saccade_Window.pdf}
%\end{center}
%\caption{Saccade window detection plot}
%\label{fig: coordinates}
%\end{figure}

In feature extraction preprocessing, a signal differentiation will first be performed for $E(t)$ as

\begin{equation}
    \frac{d}{dt} E(t) = \lim_{\Delta t \to 0} \frac{E(t) - E(t - N\Delta t)}{N\Delta t}
\end{equation}

Where $\Delta t$ represents the time difference between each sample point, and N denotes the lag sample points, which should be adjusted based on the sampling frequency. In this work, N is chosen to be 3 to capture the saccade feature better. The signal derivative $\frac{d}{dt} E(t)$ will then be used for both peak detection and saccade window detection. For peak detection, a dynamically generated threshold $s_p$ is computed based on the standard deviation of the signal derivative. This threshold is then used to identify peaks. For every detected peak, the corresponding time points $t$ are recorded in $t_{\text{p}}$. Since each peak in the signal derivative typically consists of multiple consecutive points exceeding the threshold, our algorithm groups these closely spaced points within a 500 ms window, based on the empirically tested average saccade duration. This ensures that each peak $i$ is marked only once. The process can be expressed as:

\begin{equation}
    t_{\text{p}_i} = \min \left( \left\{ t \mid \left| \frac{d}{dt} E(t) \right| \geq s_p, t \in T_i \right\} \right),
\end{equation}

where $T_i$ represents the set of time points within the 500 ms window for peak $i$, and $\min(\cdot)$ ensures that only the earliest detected point within each window is selected as $t_{\text{p}_i}$.

Since blink artifacts have already been removed and the drift frequency is significantly lower than that of saccade movements, all detected peaks can assumed to be attributed to saccades. The peak time $t_{\text{p}_i}$ will then be passed to the saccade window detector, which will search for the start and end point of each saccade event detected by the peak detector. A similar standard deviation-based threshold, $s_i$, is automatically generated for each peak to determine its start and end time. The time of the start point, $t_{p_{s_i}}$, is identified as the time where the first point lower than the threshold $s_i$ when tracing backward from the peak time $t_{\text{p}_i}$. Similarly, the end time, $t_{p_{e_i}}$, is the time of the first point lower than the threshold when moving forward from $t_{\text{p}_i}$. This can be expressed in equations as below:

\begin{equation}
    t_{p_{s_i}} = \arg \max_{t < t_{p}} \left( t \mid \left| \frac{d}{dt} E(t) \right| < s_i \right)
\end{equation}

\begin{equation}
    t_{p_{e_i}} = \arg \min_{t > t_{p}} \left( t \mid \left| \frac{d}{dt} E(t) \right| < s_i \right)
\end{equation}

The saccade start time and end time can be directly determined from $t_{p_{s_i}}$ and $t_{p_{e_i}}$, as the peak reflects the change in the EOG signal caused by the saccade. Consequently, the saccadic signal $\hat{f_s}(t)$ can be reconstructed by extracting the original EOG signal between each pair of $t_{p_{s_i}}$ and $t_{p_{e_i}}$. This can be expressed as:

\begin{equation}
    \hat{f_s}(t) =
    \begin{cases} 
        E(t), &  t \in \bigcup\limits_{i} [t_{p_{s_i}}, t_{p_{e_i}}] \\ 
        0, & \text{otherwise}
    \end{cases}
\end{equation}

The feature of saccade can now be successfully subtracted from the original EOG signal, and the saccade-excluded signal $\hat{f_{se}}(t)$ can be obtained by

\begin{equation}
    \hat{f_{se}}(t) = E(t) - \hat{f_s}(t)
\end{equation}

\subsubsection{Drift trend approximation}

At this stage, the saccade excluded signal $\hat{f_{se}}(t)$ is further adjusted by the Baseline Reconstructor, ensuring it reflects a scenario where no saccades occurred during the trial. This process involves connecting all floating segments of the EOG signal in $\hat{f_{se}}(t)$ to their preceding values and forming a continuous baseline. The baseline reconstruction process can be summarized below.

Based on the prior information of $t_{p_{s_i}}$ and $t_{p_{e_i}}$, the Baseline Reconstructor extracts the signal after each saccade at $t_{p_{e_i}}$ until it reaches either the beginning of the next saccade $t_{p_{s_{i+1}}}$ or the end of the trial $t_{end}$, whichever comes first. This can be represented as:

\begin{equation}
    f_{\text{f}_i}(t) =
    \begin{cases} 
        E(t), & t \in [t_{p_{e_i}}, \min(t_{p_{s_{i+1}}}, t_{end})] \\ 
        0, & \text{otherwise}
    \end{cases}
\end{equation}

After obtaining the floating segments $f_{\text{f}_i}(t)$, the baseline reconstructor calculates $\delta_i$ iteratively. 

The segment difference $\delta_i$ is computed as the value difference between the mean of the first $m$ samples preceding $n_{p_{s_i}}$ and the mean of the first $m$ samples following $n_{p_{e_i}}$, where $n_{p_{s_i}}$ and $n_{p_{e_i}}$ represent the start and end indices of saccade $i$ in the sampled signal, respectively. This difference is used as a displacement adjustment for the floating segment to aid in reconstructing a continuous baseline.  

More specifically, $\delta_i$ is added to the floating segment $f_{\text{f}_i}(t)$ to obtain the adjusted segment value $f_{\text{a}_i}(t)$. Notably, the first segment is adjusted based on the original $E(t)$ signal, assuming that the initial five seconds of calibration represent a drift-free true baseline. Subsequent segments are then corrected using the previously adjusted values.  

Mathematically, this process can be expressed as follows:

For \( i = 1 \):

\begin{equation}
    \delta_1 = \frac{1}{m} \sum_{k=0}^{m-1} E[n_{p_{s_1}} - k] - \frac{1}{m} \sum_{k=0}^{m-1} E[n_{p_{e_1}} + k]
\end{equation}

\begin{equation}
    f_{\text{a}_1}(t) = f_{\text{f}_1}(t) + \delta_1, \quad t \in [t_{p_{e_1}}, \min(t_{p_{s_{2}}}, t_{end})]
\end{equation}

For \( i \geq 2 \):

\begin{equation}
    \delta_i = \frac{1}{m} \sum_{k=0}^{m-1} f_{\text{a}_{i-1}}[n_{p_{s_i}} - k] - \frac{1}{m} \sum_{k=0}^{m-1} f_{\text{f}_{i}}[n_{p_{e_i}} + k]
\end{equation}

\begin{equation}
    f_{\text{a}_i}(t) = f_{\text{f}_i}(t) + \delta_i, \quad t \in [t_{p_{e_i}}, \min(t_{p_{s_{i+1}}}, t_{end})]
\end{equation}

The adjusted segment value $f_{\text{a}_i}(t)$ is then used to update the saccade-excluded signal $\hat{f_{se}}(t)$, ensuring continuity with the segment preceding the saccade. The updated saccade-excluded signal $\hat{f_{se}}(t)$ subsequently forms the reconstructed baseline $\hat{f_b}(t)$.  

\begin{equation}
    \hat{f_b}(t) =
    \begin{cases} 
        f_{\text{a}_i}(t), & t \in \bigcup\limits_{i}[t_{p_{e_i}}, \min(t_{p_{s_{i+1}}}, t_{end})] \\
        \hat{f_{se}}(t), & \text{otherwise}
    \end{cases}
\end{equation}

The obtained reconstructed EOG baseline signal $\hat{f_b}(t)$ can then be passed to the 1D multilevel wavelet decomposition mentioned for drift trend approximation. The discrete wavelet transform is applied to the reconstructed EOG baseline signal $\hat{f_b}(t)$ to decompose it and obtain approximation coefficients $A_j[n]$ and detail coefficients $D_j[n]$ at chosen level $j$.

\begin{equation}
    A_j[n] = \sum_k h[k] A_{j-1}[2n - k]
\end{equation}

\begin{equation}
    D_j[n] = \sum_k g[k] A_{j-1}[2n - k]
\end{equation}

where $h[k]$ and $g[k]$ denote low-pass filter and high-pass filter, respectively. After j levels of decomposition, the reconstructed EOG baseline signal $\hat{f_b}(t)$ can be approximated as

\begin{equation}
    \hat{f_b}(t) \approx A_j + \sum_{j=1}^{j} D_j
\end{equation}
The lowest frequency coefficient $A_j$ can then be used to reconstruct the approximation of drift trend $\hat{d}(t)$. And the de-drifted EOG signal $E'(t)$ can be obtained by subtracting the drift trend $\hat{d(t)}$ from original EOG signal $E(t)$ as

\begin{equation}
    E'(t) = E(t) - \hat{d(t)}
\end{equation}

%\begin{figure}[!t]
%\begin{center}
%\includegraphics[width=0.9\columnwidth]{Reconstuct_Base.pdf}
%\end{center}
%\caption{Plot of original EOG signal versus reconstructed baseline}
%\label{fig: coordinates}
%\end{figure}

\subsection{Gaze Estimation and Evaluation}

Once the de-drifted signal $E'(t)$ is obtained, a regression model is employed to fit $E'(t)$ to the reference signal. In this study, the reference signal is the gaze signal $\theta_{\text{r}}(t)$ obtained from the camera-based eye-tracking system PtGaze, as described in the methodology section. 

The fitted regression model’s weights are then applied to a separate de-drifted $E'(t)$ dataset to predict the gaze angle $\hat{\theta}(t)$. To assess performance, the predicted gaze angle for each saccade is evaluated by comparing the mean of the predicted angles $\overline{\hat{\theta}}_s$ to the mean of the reference angles 
$\overline{\theta}_{\text{r}_s}$ after each saccade, with the error $\epsilon(\theta_s)$ computed as:

\begin{equation}
    \epsilon(\theta_s) = \overline{\theta}_{\text{r}_s} - \overline{\hat{\theta}}_s
\end{equation}

\section{RESULTS AND DISCUSSION}

The evaluation of the proposed FGD system is divided into two parts. 

The first part focuses on testing and evaluating the FGD system against other common techniques in data simulation. These evaluations are conducted using a real-world EOG signal sample with minimal observed drift, collected using the instruments described in the methodology section. To effectively simulate various possible drift conditions, random drift noise is manually injected into the signal. The second part applies the same evaluation method from the simulation to perform gaze prediction and evaluation using real data captured from experimental trials.

The second part of the evaluation involves real data collected from five signal acquisition trials. This dataset contains 90 saccade events and is used to implement the proposed FGD system for real data de-drifting evaluation. The gaze prediction step utilizes the same regression model trained in the simulation phase.

\subsection{Evaluation Setup}

For the setup of common de-drifting methods used for comparison, a 5\textsuperscript{th}-order polynomial fitting was selected for evaluation based on performance testing with our dataset. The high-pass filtering method was evaluated using a cutoff frequency of 0.3 Hz, determined through performance testing.  

For 1D multilevel wavelet decomposition, the decomposition level $j$ was set to 7, as it provided the best approximation of the drift trend. This setting $j=7$ was applied to both the wavelet decomposition used in the baseline method evaluation and in the proposed FGD system.  

During baseline reconstruction, the segment difference $\delta_i$ was calculated using a sample count of $m=15$. This value should be adjusted accordingly based on different sampling frequencies.

To obtain the reference gaze angle for the EOG signal of this work, a camera-based gaze tracking system used with a webcam with the configuration of 720P/30HZ is used to capture the subject’s eye movement. The camera-based gaze tracking system used in this work is PtGaze with the ETH-XGaze model by X. Zhang et al. \cite{zhang2020eth}, which outperforms many other current state-of-the-art methods when testing on our dataset and suffices the goal of this work.

In each trial for data acquisition, the subject is seated in front of a screen at an approximate distance of 44 cm between the eyes and the display. To guide the subject’s gaze, a target guide consisting of nine equally spaced horizontal dots extends from 20 cm to the left to 20 cm to the right, labeled \textit{L4} to \textit{R4}, as shown in Fig. 3, while maintaining a fixed distance between each point.

\begin{figure}[!t]
\begin{center}
\includegraphics[width=1\columnwidth]{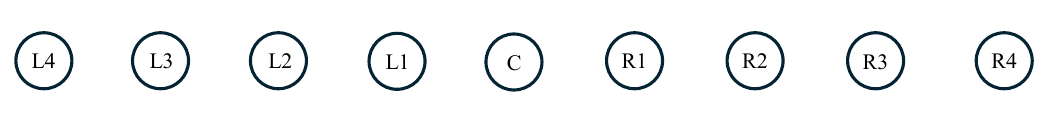}
\end{center}
\caption{Target Guide Configuration}
\label{fig: Target Guide}
\end{figure}

This range is chosen based on observations that gaze behavior remains linear within a ±30-degree range \cite{HPacuna2014eye} \cite{DIFFryu2019eog}, and the 40 cm span falls within this linear range. The guide displays only one dot at a time while the camera continuously captures and predicts the reference gaze angle. Initially, the center dot \textit{C} is presented for calibration. Subsequently, the left-side dots, \textit{L1–L4}, are displayed sequentially, with the guide returning to the center after each step. The process is then repeated for the right-side dots, \textit{R1–R4}.

\subsection{Simulation Evaluation}
In simulation evaluation, an experiment is designed to evaluate the effectiveness of the de-drifting method. As shown in Fig. 4, a real-world EOG signal sample with minimal observed drift is used, as previously mentioned, with various randomly generated low-frequency nonlinear and linear noise components injected to simulate different possible drift scenarios. The frequency of the noise signal is chosen to be $<0.1$ Hz to match the frequency range of EOG baseline drift $d(t)$. A simulation dataset with 10 different drift-simulated scenarios, containing a total of 160 saccade events, is generated through the signal noise injection process. To better illustrate the performance of different components of our system, we present the results of each processing step using a representative simulation data sample. 

\begin{figure}[!t]
\begin{center}
\includegraphics[width=0.9\columnwidth]{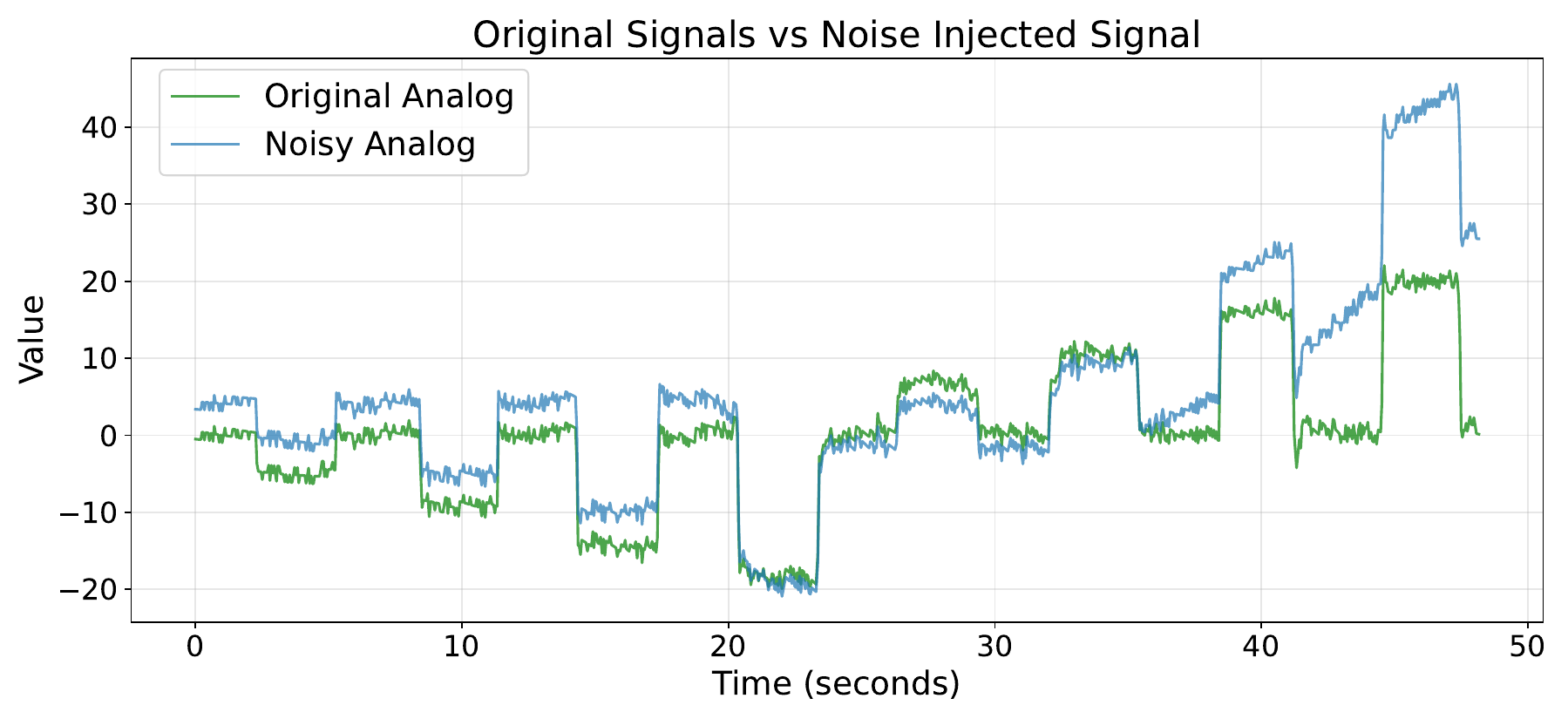}
\end{center}
\caption{Noise-injected EOG signal used for the simulation experiment. The green line represents the original EOG signal, while the blue line represents the signal after noise injection. This demonstrates a successful noise injection process that effectively mimics the drift effect in the EOG signal.}
\label{fig: NoiseInjection}
\end{figure}

Once generated, the noisy signal will be used as the raw signal $E(t)$ to test the proposed FGD system. Fig. 5 shows the sample with each peak marked by the Peak Detector, from which it can be determined that the system can successfully mark every saccade event. The Saccade Window Detector can then proceed and mark all saccade events with their start and end times. Fig. 6 shows one of the saccade windows marked by the saccade window detector within the noisy signal $E(t)$.

\begin{figure}[!t]
\begin{center}
\includegraphics[width=0.9\columnwidth]{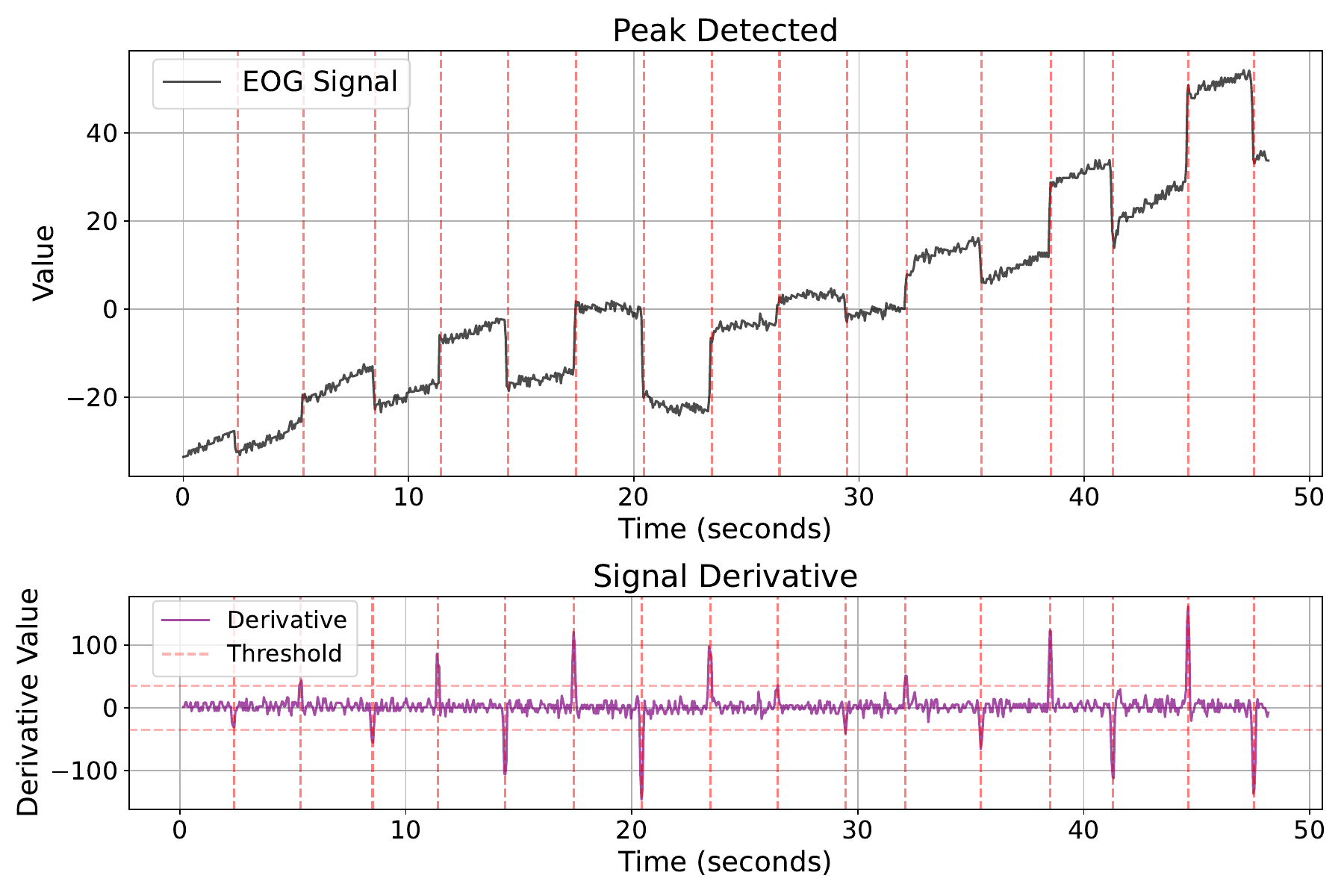}
\end{center}
\caption{Derivative plot for peak detection. The black line represents the EOG signal, while the purple line represents its derivative. The red dotted line indicates the threshold value $s_p$ and marks the detected peak positions in both signals. This demonstrates that the peak detector effectively identifies peaks, successfully marking them in both the EOG signal and its derivative.}
\label{fig: DerivativePeak}
\end{figure}

\begin{figure}[!t]
\begin{center}
\includegraphics[width=0.9\columnwidth]{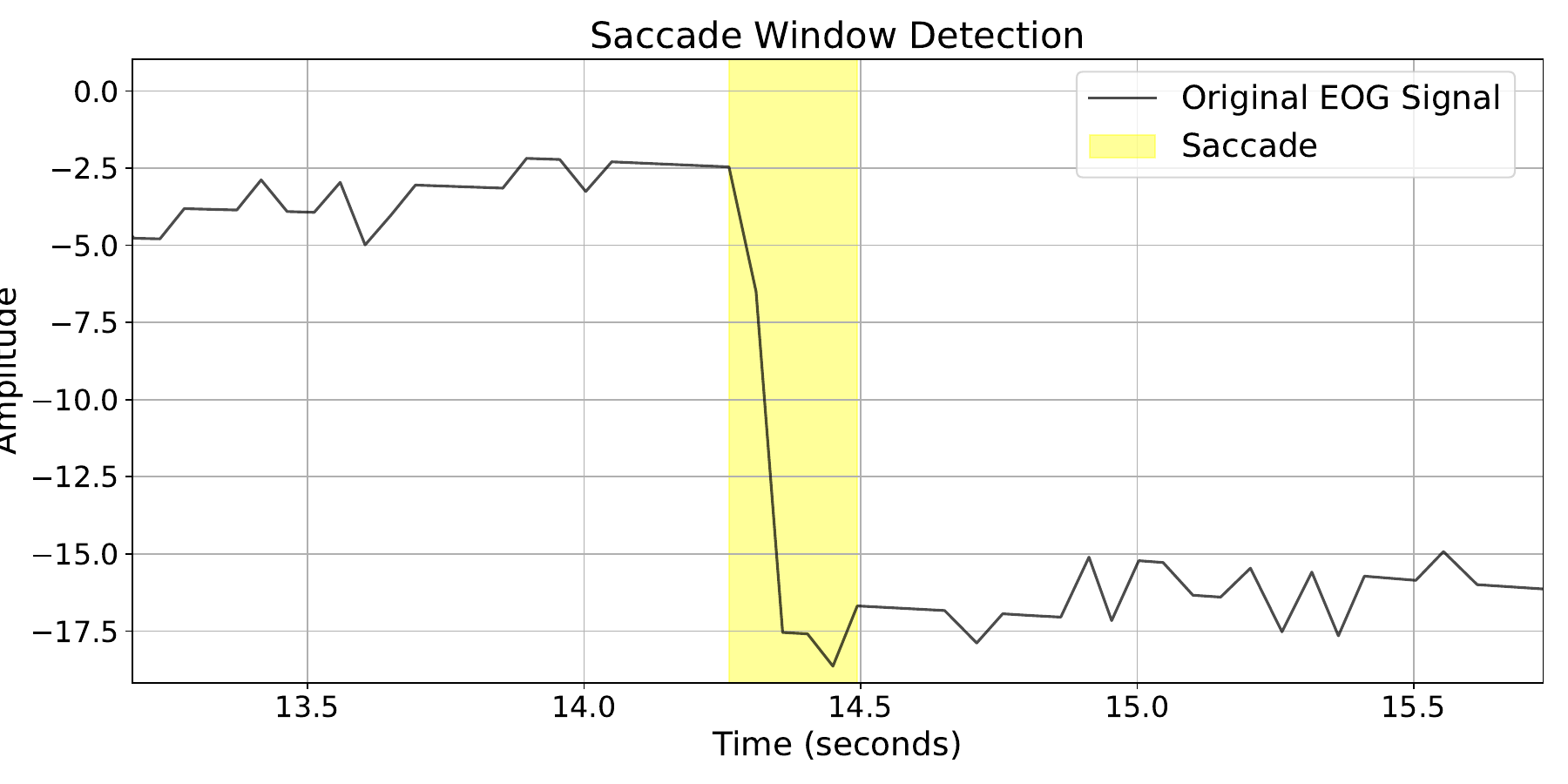}
\end{center}
\caption{Saccade window detection plot. The black line represents a segment of the EOG signal, while the yellow highlighted region indicates the saccade event detected by the Saccade Window Detector. This demonstrates that the Saccade Window Detector accurately identifies the saccade start and end times, effectively recognizing saccade events.}
\label{fig: SaccadeWindow}
\end{figure}

With the saccade feature reconstructed based on prior information, the reconstructed EOG baseline signal $\hat{f_b}(t)$ can be obtained according to Eq. (14). As illustrated in Fig. 7, the baseline reconstruction effectively aligns all floating segments to the baseline using Eqs. (9)–(14), meaning the saccade feature has been successfully extracted while preserving all other signal details. The reconstructed signal baseline achieves the objective of generating a signal as if no saccade events had occurred. Consequently, we can assume the remaining variations in the reconstructed signal $\hat{f_b}(t)$ are primarily due to the baseline drift $d(t)$. The drift trend $\hat{d(t)}$ can then be determined using a 1D multilevel wavelet decomposition at level 7, and its trend approximation can be found in Fig. 8. The sample will then complete its de-drifting using the trend and regression fitting will be used to obtain the gaze prediction from this sample dataset.
\begin{figure}[!t]
\begin{center}
\includegraphics[width=0.9\columnwidth]{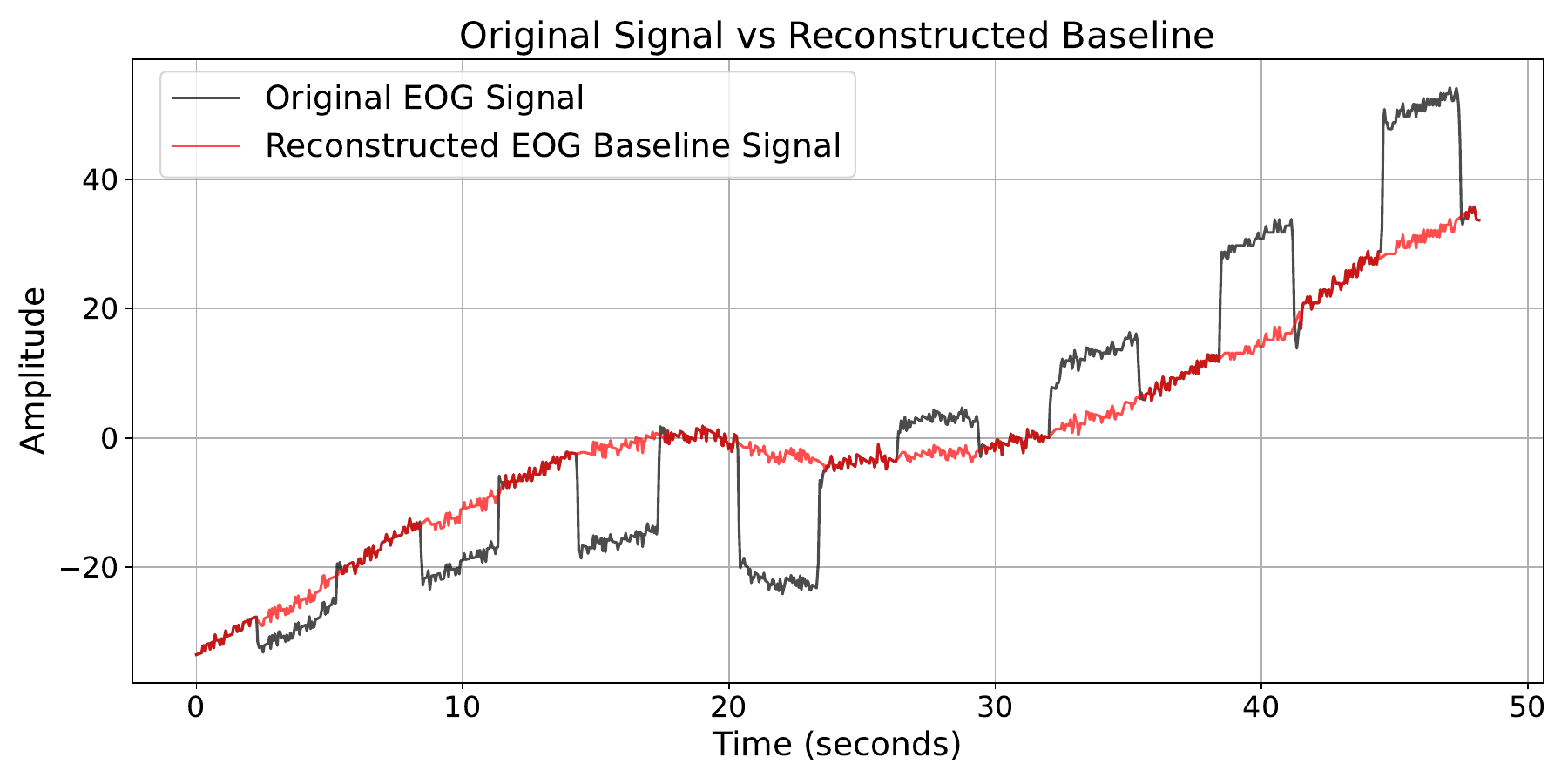}
\end{center}
\caption{Plot of the original EOG signal versus the reconstructed baseline. The black line represents the EOG signal, while the red line denotes the reconstructed EOG baseline signal $\hat{f_b}(t)$. This demonstrates that the Baseline Reconstructor successfully creates a continuous baseline for the EOG signal by adjusting the floating segments in the saccade-excluded signal $\hat{f_{se}}(t)$.}
\label{fig: ReconstructedBase}
\end{figure}

\begin{figure}[!t]
\begin{center}
\includegraphics[width=0.9\columnwidth]{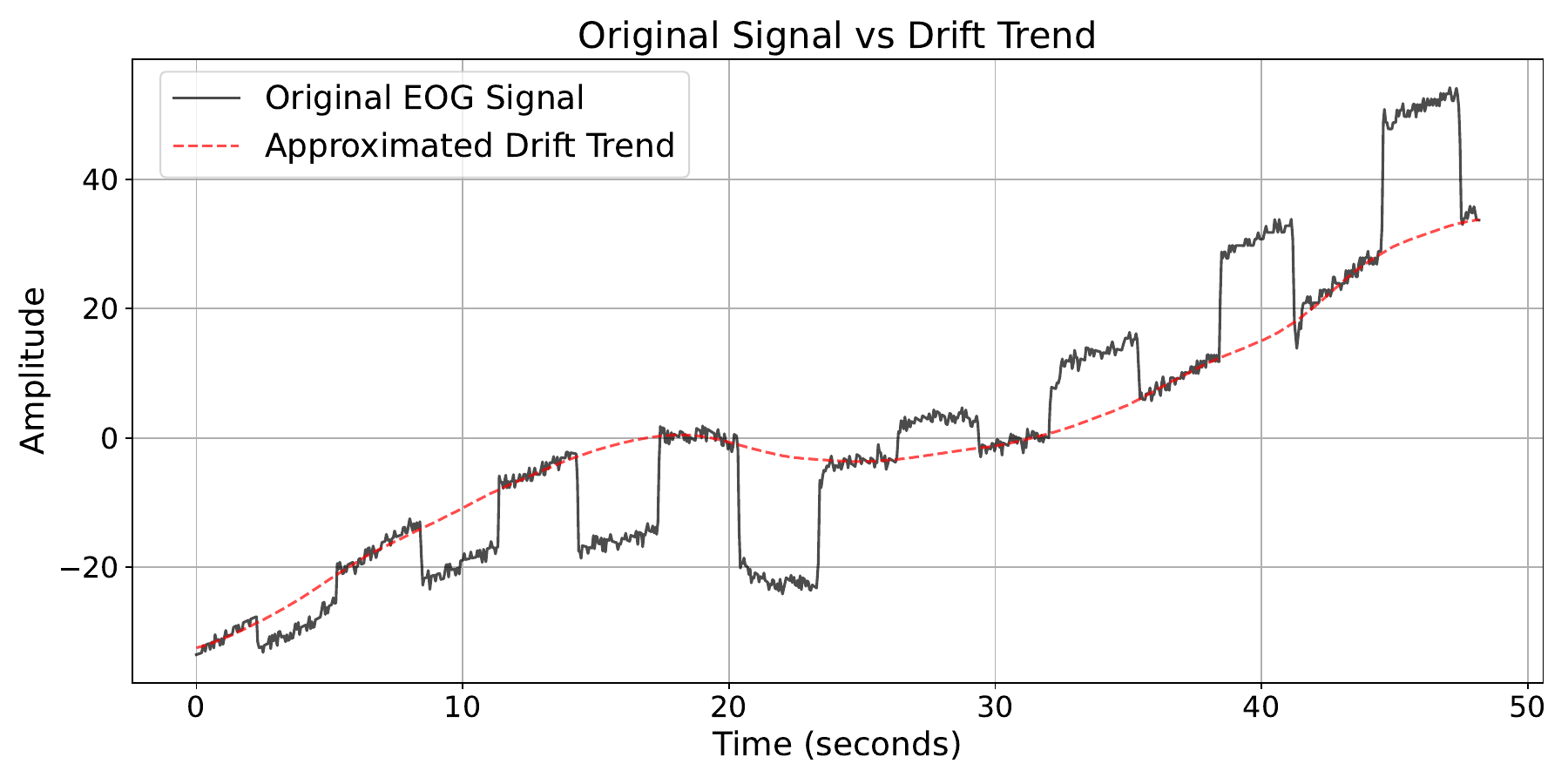}
\end{center}
\caption{Drift trend approximation for the original EOG signal. The black line represents the EOG signal, while the red dotted line represents the approximated drift trend. This demonstrates that 1D Wavelet Decomposition can effectively utilize the reconstructed EOG baseline signal $\hat{f_b}(t)$ to accurately approximate the drift trend of the original EOG signal.}
\label{fig: Drift_Trend}
\end{figure}

\subsection{Comparison and evaluation}

In this section, other de-drifting techniques are also implemented for comparison and evaluation against our results, utilizing the same regression model and settings. In Fig. 9, a simulation sample illustrates the de-drifting outcomes from four methods: (a) polynomial fitting, (b) high-pass filtering, (c) 1D multilevel wavelet decomposition, and (d) the proposed FGD system. From the results, we infer that both high-pass filtering and 1D multilevel wavelet decomposition distort the original signal during the de-drifting process, altering its morphology by affecting the amplitude of key eye movement features such as saccades. In contrast, polynomial fitting preserves the signal morphology well but is less effective in de-drifting compared to the proposed FGD system.

\begin{figure}[!t]
    \centering
    \begin{subfigure}{0.45\textwidth}
        \centering
        \includegraphics[width=0.9\columnwidth]{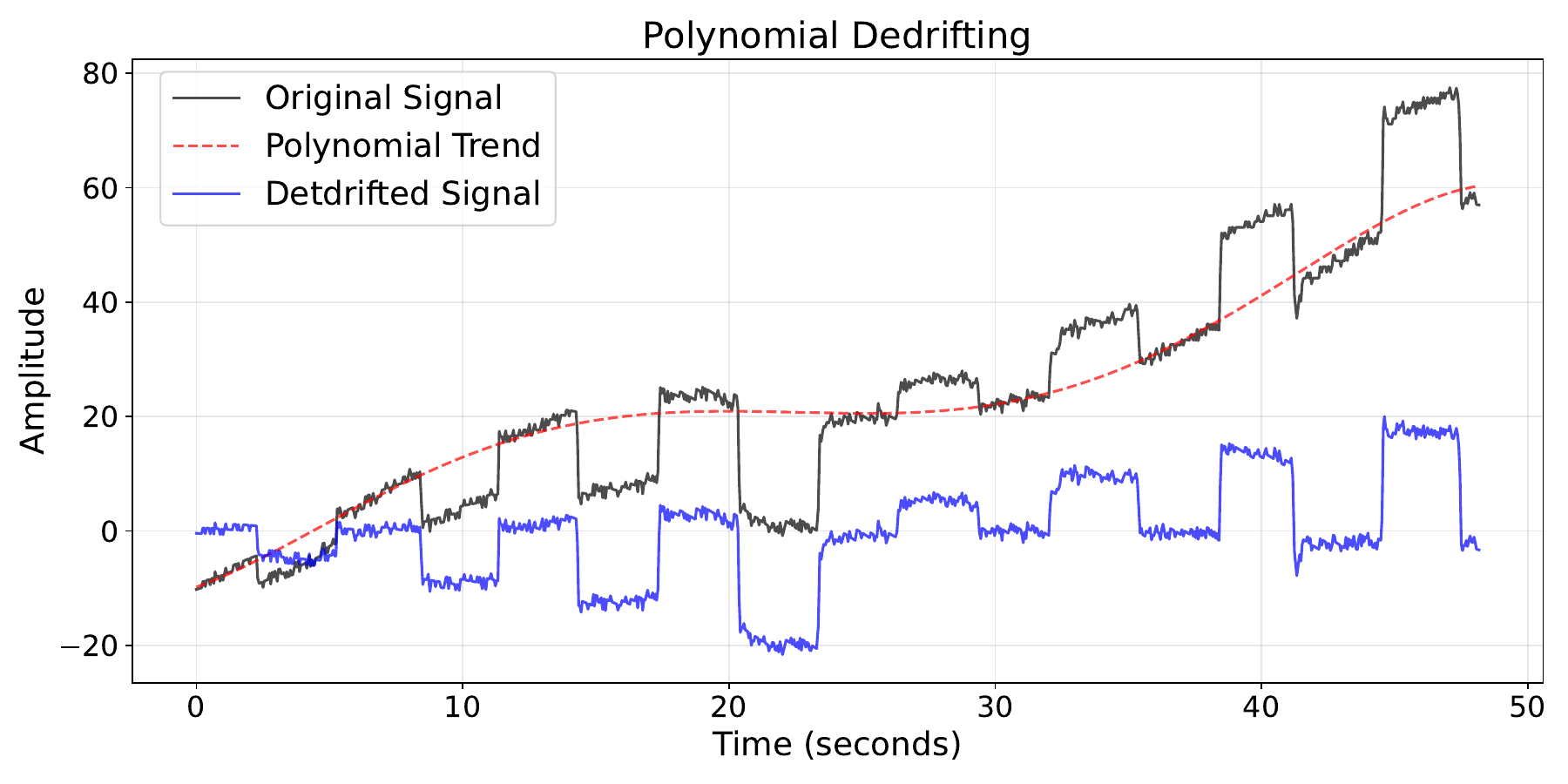}
        \caption{Polynomial fitting de-drifting}
        \label{fig: coordinates}
    \end{subfigure}
    \begin{subfigure}{0.45\textwidth}
        \centering
        \includegraphics[width=0.9\columnwidth]{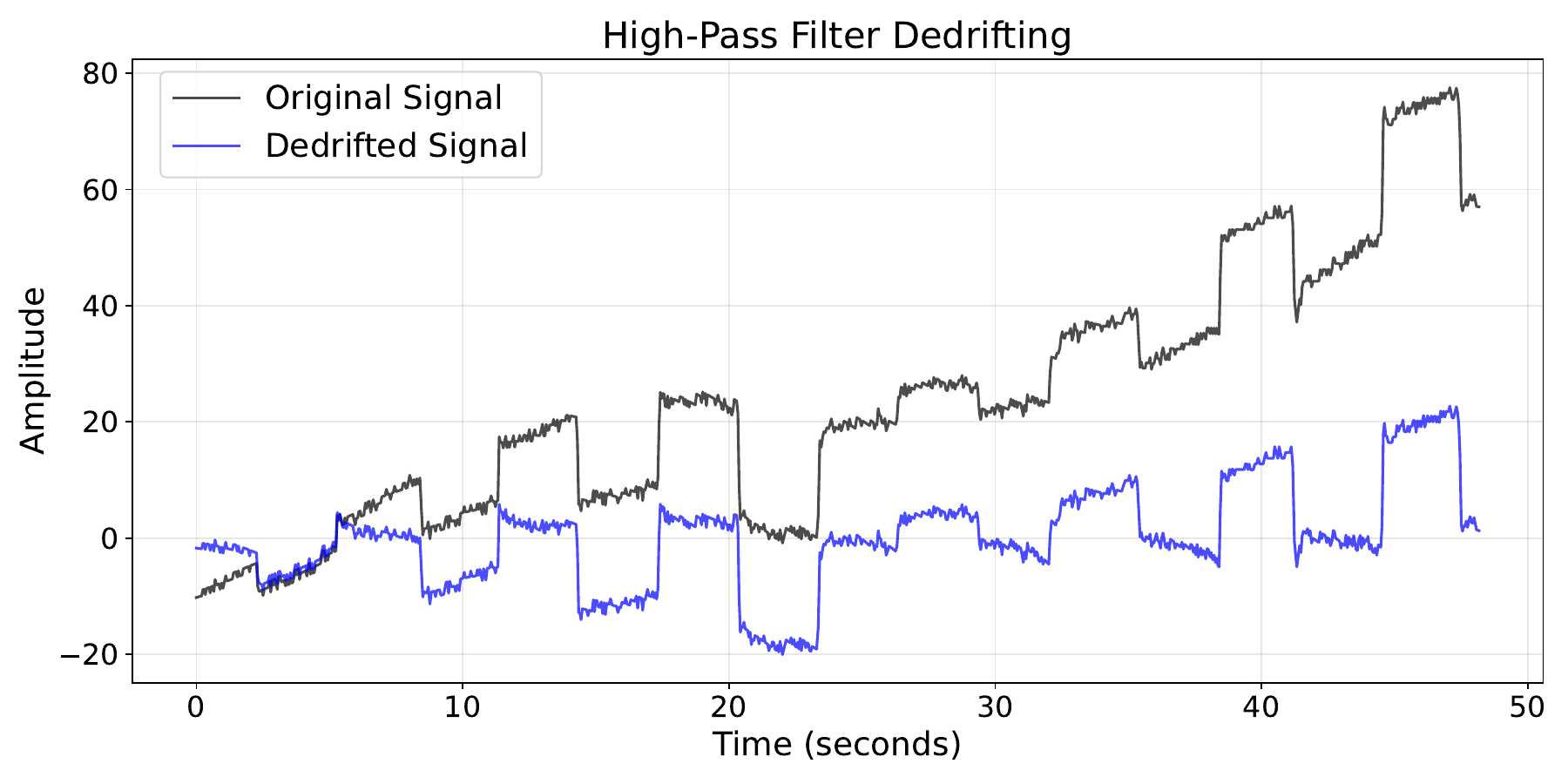}
        \caption{High-pass filtering de-drifting}
        \label{fig:fig2}
    \end{subfigure}
    
    \begin{subfigure}{0.45\textwidth}
        \centering
        \includegraphics[width=0.9\columnwidth]{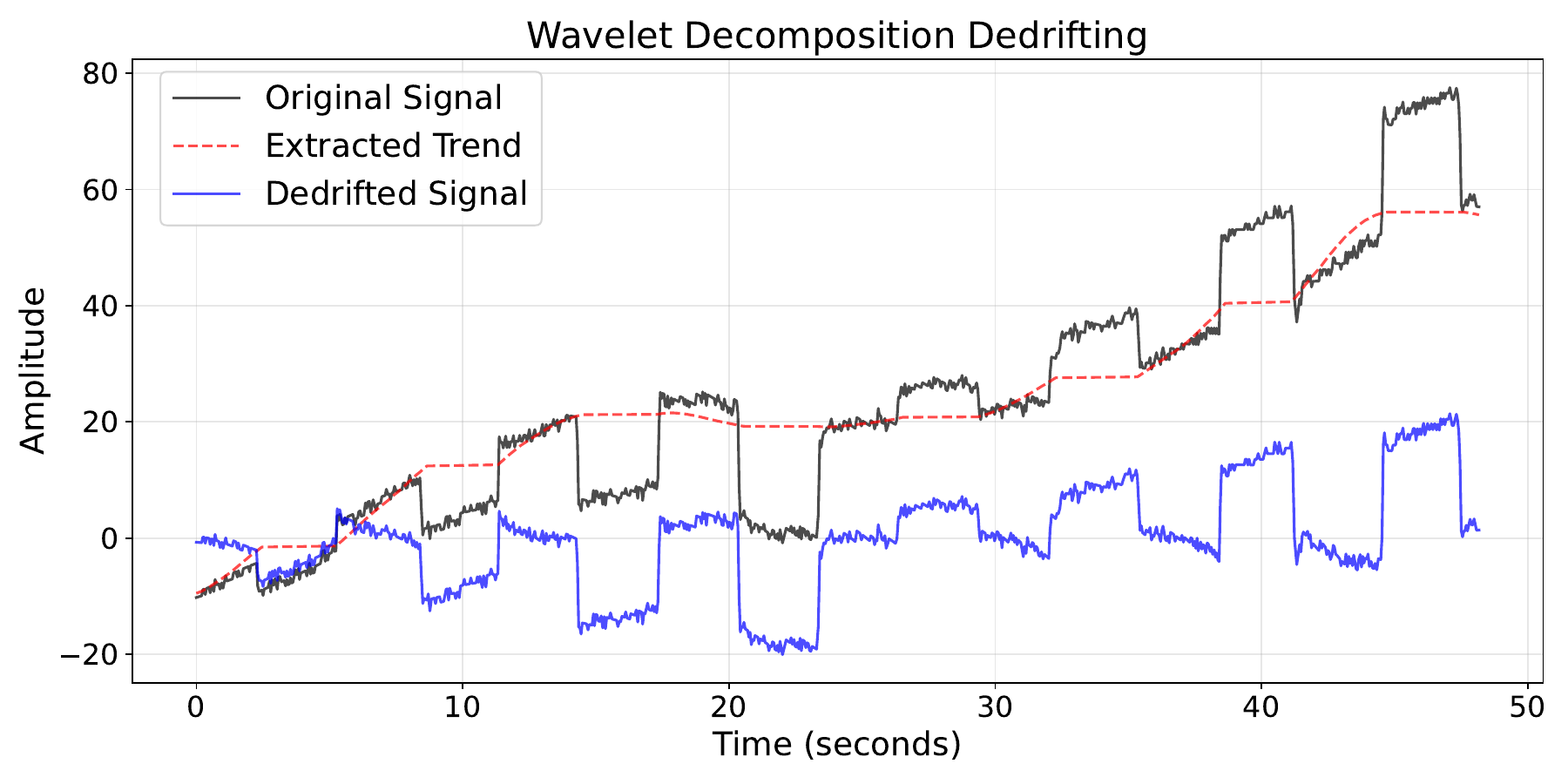}
        \caption{Wavelet decomposition de-drifting}
        \label{fig:fig3}
    \end{subfigure}
    \begin{subfigure}{0.45\textwidth}
        \centering
        \includegraphics[width=0.9\columnwidth]{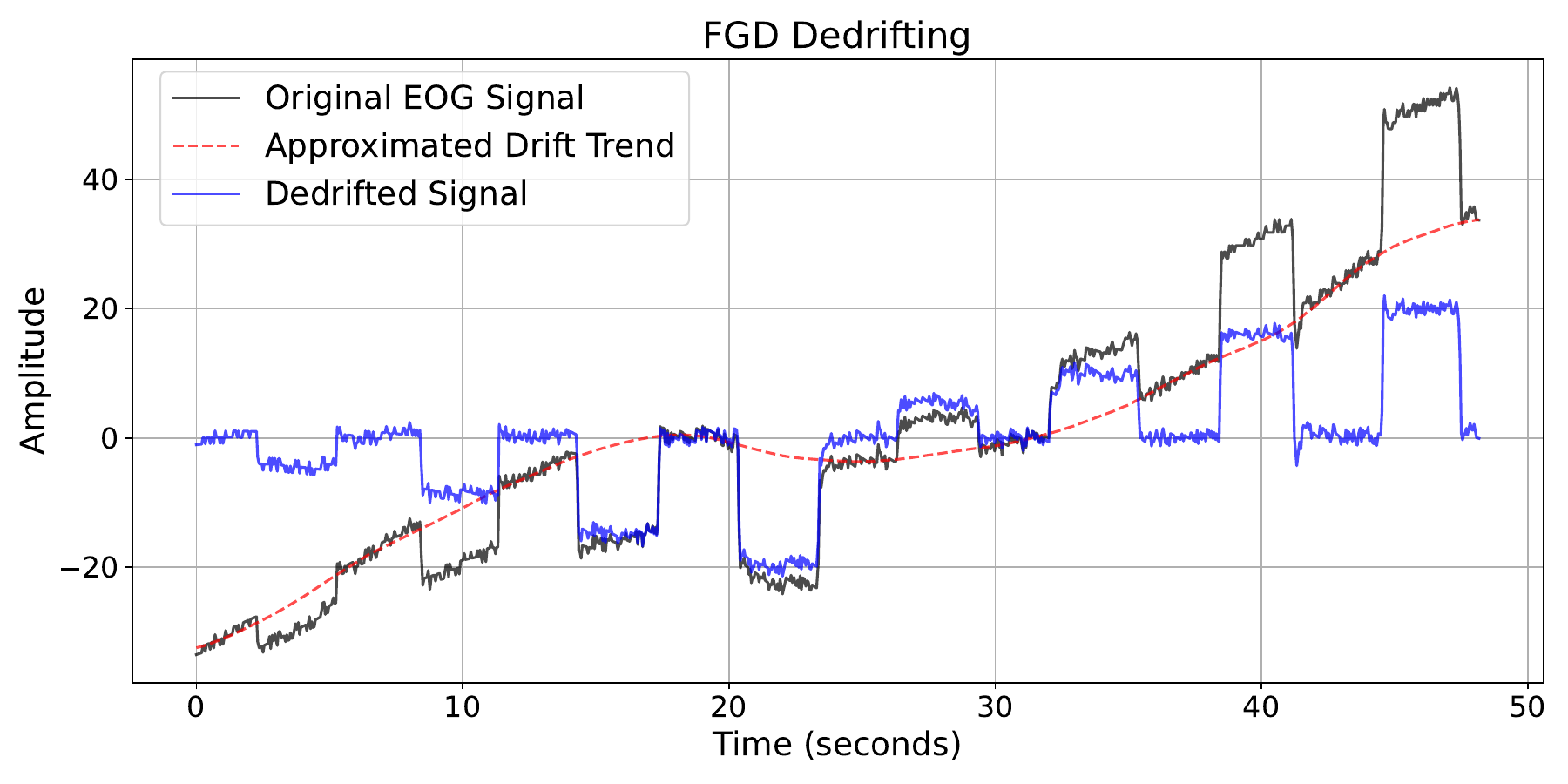}
        \caption{Feature-Guided De-drifting}
        \label{fig:fig4}
    \end{subfigure}

    \caption{De-drifting results with different methods. The black line represents the original EOG signal, the blue line represents the de-drifted EOG signal, and the red dotted line represents the drift trend approximated by each method. This demonstrates a better performance of the proposed FGD system compared to other methods.}
    \label{fig:combined}
\end{figure}

\begin{table*}[thpb!]
\caption{De-Drifting Performance of the Proposed FGD Compared to Other Methods Based on Gaze Prediction Error}
\label{table_example}
\begin{center}
\begin{tabular}{|c|c|c|c|c||c|}
\hline
\multicolumn{5}{|c||}{Simulation} & Real Data \\ % Corrected multicolumn alignment
\hline
Target & \makecell{Proposed FGD \\ $\epsilon(\theta_s)$ (°)} & \makecell{Polynomial Fitting \\ $\epsilon(\theta_s)$ (°)} & \makecell{High-pass Filtering \\ $\epsilon(\theta_s)$ (°)} & \makecell{Wavelet Decomposition \\ $\epsilon(\theta_s)$ (°)} & \makecell{Proposed FGD \\ $\epsilon(\theta_s)$ (°)} \\
\hline
L4 & 0.548 & 0.980 & 1.585 & 1.389 & 0.983\\
\hline
L3 & 0.959 & 2.851 & 2.803 & 1.753 & 1.658\\
\hline
L2 & 1.469 & 2.029 & 2.143 & 2.055 & 1.040\\
\hline
L1 & 1.009 & 1.742 & 2.438 & 1.298 & 0.523\\
\hline
R1 & 0.521 & 1.078 & 2.148 & 0.518 & 0.543\\
\hline
R2 & 1.096 & 1.518 & 2.442 & 1.828 & 1.253\\
\hline
R3 & 0.911 & 1.873 & 2.631 & 0.842 & 1.145\\
\hline
R4 & 0.658 & 1.793 & 1.886 & 1.569 & 1.123\\
\hline
Average & \textbf{0.896 ± 0.297} & 1.733 ± 0.586 & 2.260 ± 0.400 & 1.406 ± 0.516 & \textbf{1.033 ± 0.371}\\
\hline
\end{tabular}
\end{center}
\end{table*}

The simulation gaze prediction results for each de-drifted dataset, processed using the FGD system and other common de-drifting methods, are presented in Table I, along with the FGD prediction results on real-world data.

As shown in Table I, the error produced by the proposed FGD method in both simulation and real data is smaller than that of other evaluated methods in the simulation. This demonstrates the superiority of the proposed FGD method, achieving a 36.29\% reduction in mean error in simulation and a 26.53\% reduction in real data, compared to the best-performing alternative (Wavelet Decomposition). Additionally, the error using real data is slightly higher than that observed with simulated data, probably due to the simulation not capturing all possible drift characteristics, and real data may include additional noise and artifacts not present in the simulation.  

However, despite the presence of more unpredictable noise and drift variations in real data, the proposed method still achieves higher accuracy and outperforms other approaches evaluated in the simulation, further validating its effectiveness. With our FGD system, the EOG signal captured in a free-movement HRC scenario can be more effectively de-drifted, resulting in improved eye gaze prediction for further analysis.

Future work will focus on adapting the proposed method for real-time applications. This includes optimizing the feature extraction process for real-time performance and replacing the current offline 1D multilevel wavelet decomposition with low-latency detrending techniques such as Kalman filtering. Future efforts will also assess the computational cost and latency of each module to ensure feasibility and minimize the delay introduced for real-time scenarios like closed-loop robotic control. Additionally, the current dataset is still limited in size and diversity, and expanding the dataset will be a priority to enhance model generalization and robustness.

\section{CONCLUSION}
   
To support long-term EOG-based eye tracking in HRC applications, where sensor fusion requires a drift-free signal, a more effective method is needed to mitigate drift effects. In this study, we propose and evaluate a novel Feature-Guided De-drifting (FGD) system for EOG signals, which includes active eye movement feature extraction and adaptive baseline reconstruction to remove baseline drift. The method was tested and evaluated, demonstrating a 36.29\% reduction in mean error in simulation and a 26.53\% reduction in real data, outperforming the best alternative method. The FGD system offers a more accurate and reliable solution for EOG-based gaze estimation, making it particularly suitable for enhancing human performance in HRC scenarios where the user’s gaze angle and transitions occur freely.

\addtolength{\textheight}{-3cm}   % This command serves to balance the column lengths
                                  % on the last page of the document manually  It shortens
                                  % so it should come on the page before the last  Make

%%%%%%%%%%%%%%%%%%%%%%%%%%%%%%%%%%%%%%%%%%%%%%%%%%%%%%%%%%%%%%%%%%%%%%%%%%%%%%%%

%%%%%%%%%%%%%%%%%%%%%%%%%%%%%%%%%%%%%%%%%%%%%%%%%%%%%%%%%%%%%%%%%%%%%%%%%%%%%%%%

%%%%%%%%%%%%%%%%%%%%%%%%%%%%%%%%%%%%%%%%%%%%%%%%%%%%%%%%%%%%%%%%%%%%%%%%%%%%%%%%
\bibliographystyle{IEEEtran}
\bibliography{IEEEabrv,references}

%\begin{thebibliography}{99}

%\end{thebibliography}
\end{document}